\documentstyle[aps,prl,graphicx,amsbsy]{revtex}
\begin{document}
\draft
\wideabs{
\title{Chiral fluctuations in MnSi above the Curie temperature}
\author{ B.~Roessli$^1$, P.~B\"oni$^2$, W.~E.~Fischer$^1$ and Y.~Endoh$^3$}
\address
{ $^1$Laboratory for Neutron Scattering, ETH Zurich \& Paul
Scherrer Institute, CH-5232 Villigen PSI\\ 
$^2$ Physik-Department
E21, Technische Universit\"at M\"unchen, D-85747 Garching, Germany\\
$^3$Physics Department, Tohoku University, Sendai 980, Japan
}

\date{\today}
\newpage

\maketitle

\begin{abstract}

\indent

Polarized neutrons are used to determine the antisymmetric part of
the magnetic susceptibility in non-centrosymmetric MnSi. The
paramagnetic fluctuations are found to be incommensurate 
with the chemical lattice and to have a chiral character.
 We argue that antisymmetric interactions 
must be taken into account to properly describe the critical 
dynamics in MnSi above T$_C$. 
The possibility of directly measuring the
polarization dependent part of the dynamical susceptibility in a large class
of compounds through polarised inelastic neutron-scattering is outlined as it can yield
direct evidence for antisymmetric interactions like spin-orbit
coupling in metals as well as in insulators.

\end{abstract}

 }

Ordered states with helical arrangement of the magnetic moments
are described by a chiral order parameter $\vec C=\vec S_1 \times
\vec S_2$, which yields the left- or right-handed rotation of
neighboring spins along the pitch of the helix. Examples for
compounds of that sort are rare-earth metals like Ho
\cite{plakhty_01}. Spins on a frustrated lattice form another
class of systems, where simultaneous ordering of chiral and spin
parameters can be found. For example, in the triangular lattice
with antiferromagnetic nearest neighbor
interaction, the classical ground-state is given by a
non-collinear arrangement with the spin vectors forming a
120$^\circ$ structure. In this case, the ground state is highly
degenerate as a continuous rotation of the spins in the hexagonal
plane leaves the energy of the system unchanged. In addition, it
is possible to obtain two equivalent ground states which differ
only by the sense of rotation (left or right) of the magnetic
moments from sub-lattice to sub-lattice, hence yielding an example
of chiral degeneracy.

As a consequence of the chiral symmetry of the order parameter, a
new universality class results that is characterized by novel
critical exponents
as calculated by Monte-Carlo simulations \cite{kawamura_88} and
measured by neutron scattering \cite{mason_89} in the
$XY$-antiferromagnet CsMnBr$_3$. An interesting but still
unresolved problem is the characterization of chiral spin
fluctuations that have been suggested to play an important role
e.g. in the doped high-$T_c$ superconductors \cite{sulewski}. The
measurement of chiral fluctuations is, however, a difficult task
and can usually only be performed by projecting the magnetic
fluctuations on a field-induced magnetization
\cite{maleyev95,plakhty99}.

In this Letter, we show that chiral fluctuations can be directly
observed in non-centrosymmetric crystals without disturbing the
sample by a magnetic field. We present results of polarized
inelastic neutron scattering experiments performed in the
paramagnetic phase of the itinerant ferromagnet MnSi that confirm
the chiral character of the spin fluctuations due to spin-orbit
coupling and discuss the experimental results in the framework of
self-consistent renormalisation theory of spin-fluctuations in
itinerant magnets \cite{moriya85}.

Being a prototype of a weak itinerant ferromagnet, the magnetic
fluctuations in MnSi have been investigated in the past in detail
by means of unpolarized and polarized neutron scattering. The
results demonstrate the itinerant nature of the spin fluctuations
\cite{ishikawa77,ishikawa82,ishikawa85} as well as the occurrence
of spiral correlations \cite{shirane83} and strong longitudinal
fluctuations \cite{tixier97}.

\begin{figure}
\centering
\includegraphics*[scale=0.35, angle=-90]{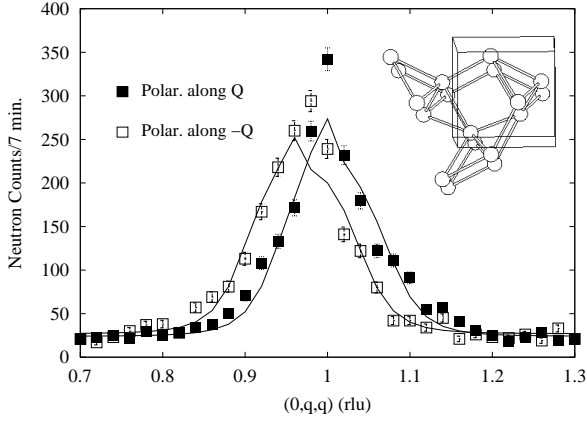}
\vskip 4pt \caption{Inelastic spectra in MnSi ($\hbar \omega =
0.5$ meV) at $T = 31$ K for the neutron polarization parallel and
anti-parallel to the scattering vector $\vec Q$, respectively. The
solid lines are fits to the data. The inset shows the Mn atoms in the crystal
structure of MnSi. Note that MnSi is not centro-symmetric. }
\label{Fig1}
\end{figure}

MnSi has a cubic space group P2$_1$3 with a lattice constant $a =
4.558$ \AA\ that lacks a center of symmetry leading to a
ferromagnetic spiral along the [1 1 1] direction with a period of
approximately 180 \AA \cite{bloch75}. The Curie temperature is
$T_C = 29.5$ K. The spontaneous magnetic moment of Mn $\mu_s
\simeq 0.4 \mu_B$ is strongly reduced from its free ion value
$\mu_f = 2.5\mu_B$. As shown in the inset of Fig.~\ref{Fig1} the
four Mn and Si atoms are placed at the positions $(x,x,x)$,
$({1\over2}+x,{1\over2}-x,-x)$, $({1\over2}-x,-x,{1\over2}+x)$,
and $(-x,{1\over2}+x,{1\over2}-x)$ with $x_{Mn} = 0.138$ and $x_{Si}=0.845$, 
resepctively.

We investigated the paramagnetic fluctuations in a large single
crystal of MnSi (mosaic $\eta = 1.5^0$) of about 10 cm$^3$ on the
triple-axis spectrometer TASP at the neutron spallation source
SINQ using a polarized neutron beam. The single crystal was mounted in a $^4$He refrigerator of
ILL-type and aligned with the [0 0 1] and [1 1 0] crystallographic
directions in the scattering plane. Most constant energy-scans
were performed around the (0 1 1) Bragg peak and in the
paramagnetic phase in order to relax the problem of depolarization
of the neutron beam in the ordered phase. The spectrometer was
operated in the constant final energy mode with a neutron wave
vector $\vec k_f$=1.97 $\AA^{-1}$. In order to suppress
contamination by higher order neutrons a pyrolytic graphite filter
was installed in the scattered beam. The incident neutrons were
polarized by means of a remanent \cite{remanent} FeCoV/TiN-type
bender that was inserted after the monochromator
\cite{semadeni01}. The polarization of the neutron beam at the
sample position was maintained by a guide field $B_g = 10$ G that
defines also the polarization of the neutrons $\vec P_i$ with
respect to the scattering vector $\vec Q = \vec k_i - \vec k_f$ at
the sample position.

In contrast to previous experiments, where the polarization $\vec
P_f$ of the scattered neutrons was also measured in order to
distinguish between longitudinal and transverse fluctuations
\cite{tixier97}, we did not analyze $\vec P_f$, as our goal was to
detect a polarization dependent scattering that is proportional to
$\sigma_p \propto (\hat{\vec Q} \cdot \vec P_i)$ as discussed
below.

A typical constant-energy scan with $\hbar \omega = 0.5$ meV
measured in the paramagnetic phase at $T = 31$~K is shown in
Fig.~\ref{Fig1} for the polarization of the incident neutrons
$\vec P_i$ parallel and anti-parallel to the scattering vector
$\vec Q$. It is clearly seen that the peak positions depend on
$\vec P_i$ and appear at the incommensurate positions $\vec Q =
\vec \tau \pm \vec\delta$ with respect to the reciprocal lattice
vector $\vec\tau_{011}$ of the nuclear unit cell. Obviously, this
shift of the peaks with respect to (0 1 1) would be hardly visible
with unpolarized neutrons and could not observed in previous inelastic 
neutron works.

In order to discuss our results we start with the general
expression for the cross-section of magnetic scattering with
polarized neutrons\cite{izyumov}
\begin{eqnarray}
{d^2\sigma\over{d\Omega d\omega}} &\sim&  \sum_{\alpha, \beta}(\delta_{\alpha, \beta}-
\hat Q_\alpha \hat Q_\beta) A^{\alpha \beta} (\vec Q, \omega) \nonumber
\\
&+&  \sum_{\alpha, \beta} (\hat {\vec Q} \cdot \vec
P_i)\sum_{\gamma}\epsilon_{\alpha, \beta, \gamma} \hat Q^\gamma
B^{\alpha \beta}(\vec Q, \omega) \label{ncs}
\end{eqnarray}
where $(\vec Q, \omega)$ are the momentum and energy-transfers
from the neutron to the sample, $\hat {\vec Q} = \vec Q/|\hat Q|$,
and $\alpha, \beta, \gamma$ indicate Cartesian coordinates. The
first term in Eq.~\ref{ncs} is independent of the polarization of
the incident neutrons, while the second is polarization dependent
through the factor $(\hat{\vec Q} \cdot \vec P_i)$. $\vec P_i$
denotes the direction of the neutron polarization and its scalar
is equal to 1 when the beam is fully polarized. $A^{\alpha \beta}$
and $B^{\alpha \beta}$ are the symmetric and antisymmetric parts
of the scattering function $S^{\alpha \beta}$, that is $A^{\alpha
\beta}={1\over 2} (S^{\alpha \beta} + S^{ \beta \alpha})$ and
$B^{\alpha \beta}={1\over 2} (S^{\alpha \beta} - S^{\beta
\alpha})$. $S^{\alpha \beta}$ are the Fourier transforms of the
spin correlation function $<s^\alpha_l s^\beta_{l'}>$, $S^{\alpha
\beta}(\vec Q, \omega)={1\over{2\pi N}}\int_{-\infty}^\infty{dt
e^{-i\omega t} \sum_{ll'}{e^{i\vec Q (\vec X_l-\vec X_{l'})
}}<s^\alpha_l s^\beta_{l'}(t)> }$. The vectors $\vec X_l$
designate the positions of the scattering centers in the lattice.
The correlation function is related to the dynamical
susceptibility through the fluctuation-dissipation theorem $S(\vec
Q,\omega)=2\hbar/(1-\exp(-\hbar\omega/kT))\Im \chi(\vec
Q,\omega)$.

Following Ref. \cite{lovesey} we define now an axial vector $\vec
B$ by 
$\sum_{\alpha \beta}\epsilon_{\alpha \beta \gamma}B^{\alpha \beta}
= B^\gamma (\vec Q, \omega)$, that 
represents the antisymmetric part of the susceptibility which, 
hence, 
depends on the neutron polarization as follows
\begin{equation}
(\hat {\vec Q }\cdot \vec P_i)(\hat {\vec Q} \cdot \vec B)
\label{axial}
\end{equation}
and vanishes for centro-symmetric systems 
 or when
there is no long-range order. In the absence of symmetry breaking
fields like external magnetic fields, pressure etc., similar scans
with polarized neutrons would yield a peak of diffuse scattering
at the zone center and no scattering that depends on the
polarization of the neutrons. However, an intrinsic anisotropy of
the spin Hamiltonian in a system that lacks lattice inversion
symmetry may provide an axial interaction leading to a
polarization dependent cross section. The polarization dependent
scattering obtained in the present experiments is therefore an indication
of fluctuations in the chiral order parameter and points towards
the existence of an axial vector $\vec B$ that is not necessarily
commensurate with the lattice. Hence, according to Eq.~\ref{axial}
the neutron scattering function in MnSi contains a non-vanishing
antisymmetric part.

Because the crystal structure of MnSi is non-centrosymmetric and
the magnetic ground-state forms a helix with spins perpendicular
to the [1 1 1] crystallographic direction, it is reasonable to
interpret the polarization-dependent transverse part of the
dynamical susceptibility in terms of the Dzyaloshinskii-Moriya
(DM) interaction \cite{dzyal58,moriya60} similarly as it was done
in other non-centrosymmetric systems that show incommensurate
ordering \cite{zheludev,roessli}.

Usually the DM-interaction is written as the cross product of
interacting spins $H_{DM}=\sum_{l,m}\vec D_{l,m}\cdot (\vec s_l
\times \vec s_m)$, where the direction of the DM-vector $\vec D$
is determined by bond symmetry and its scalar by the strength of
the spin-orbit coupling \cite{moriya60}. Although the
DM-interaction was originally introduced on microscopic grounds
for ionic crystals, it was shown that antisymmetric spin
interactions are also present in metals with non-centrosymmetric
crystal symmetry \cite{kataoka_84}. In a similar way as  for
insulators with localized spin densities, the antisymmetric
interaction originates from the spin-orbit coupling in the absence
of an inversion center and a finite contribution to the the
antisymmetric part of the wave-vector dependent dynamical
susceptibility is obtained.

\begin{figure}
\centering
\includegraphics*[scale=0.32, angle=-90]{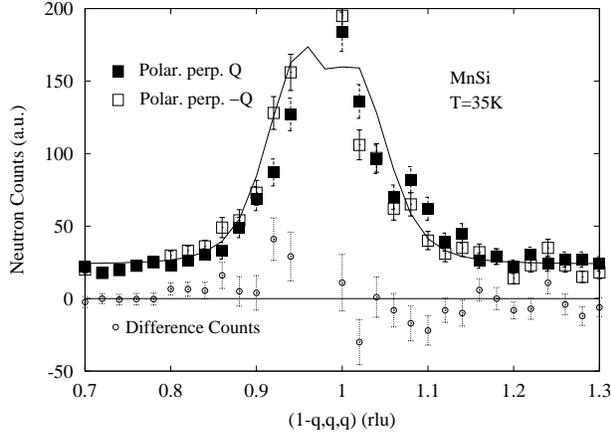}
\vskip 4pt \caption{Neutron spectra in MnSi for an energy-transfer
$\hbar \omega = 0.5$ meV as measured at $T = 35$ K for $\vec P_i$
perpendicular to $\vec Q$ and $-\vec Q$, respectively. The solid
line shows a fit to the data and the small symbols represent the
difference signal that is independent of $\vec P_i$. See text for
details.} \label{Fig2}
\end{figure}

\begin{figure}
\centering
\includegraphics*[scale=0.32, angle=-90]{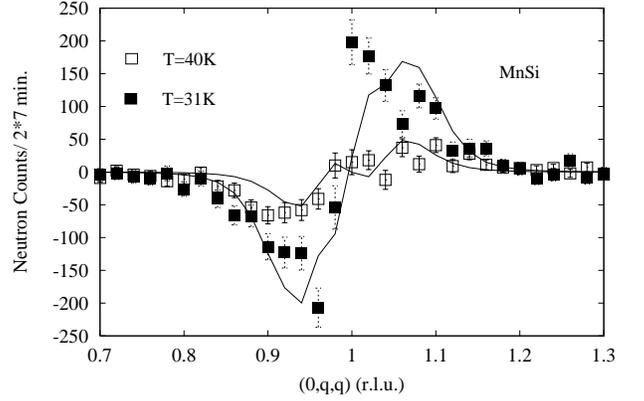}
\caption{Difference neutron counts for polarization $\vec P_i$ of
the incident neutron beam parallel and anti-parallel to $\vec Q$
in MnSi at $T = 31$ K and 40 K, respectively. The solid lines are fit 
to the data using the SRC-result for the dynamical susceptibility 
with the parameters given in the text.}
\label{Fig3}
\end{figure}

For the case of a uniform DM-interaction, the neutron cross-section depends on the
polarization of the neutron beam \cite{aristov_00} as follows
\begin{eqnarray}
 \biggl({{d^2\sigma}\over{d\Omega d\omega}}\biggr)_{np}
   & \sim & \Im{{(\chi^\perp(\vec q-\vec\delta,\omega)+\chi^\perp (\vec q+\vec\delta,\omega))} }, \nonumber \\
 \biggl({{d^2\sigma}\over{d\Omega d\omega}}\biggr)_{p}
   & \sim & (\hat{\vec D} \cdot \hat{\vec Q})(\hat{\vec Q}\cdot \vec P_i) \nonumber \\
   & \times & \Im{{(\chi^\perp(\vec q-\vec\delta,\omega)-\chi^\perp (\vec q+\vec\delta,\omega))}
   }.
\label{pa}
\end{eqnarray}
Here, $\vec q$ designates the reduced momentum transfer with
respect to the nearest magnetic Bragg peak at $\vec \tau \pm \vec
\delta$. The first line of Eq. \ref{pa} describes inelastic
scattering with a non-polarized neutron beam. The second part
describes inelastic scattering that depends on $\vec P_i$ as well
as on $\vec D$. Eq. \ref{pa} shows that the cross section for
$\vec P_i \perp \vec Q$ is indeed independent of $P_i$ as observed
in Fig. \ref{Fig2}. By subtracting the inelastic spectra taken
with $\vec P_i$ parallel and anti-parallel to $\vec Q$, the
polarization dependent part of the cross-section can be isolated,
as demonstrated in Fig. \ref{Fig3} for two temperatures $T = 31$ K
and $T = 40$ K.

Close to $T_C$, the intensity is rather high and the crossing at
$Q =$ (0 1 1) is sharp. At 40 K the intensity becomes small and
the transition at (0 1 1) is rather smooth, which mirrors the
decreases  of the correlation length with increasing temperature.
We have measured $({{d^2\sigma}/({d\Omega d\omega}}))_{p}$ in the
vicinity of the (0 1 1) Bragg peak at $T = 35$ K. The result shown
as a contour plot in Fig. \ref{Fig4} indicates that the
DM-interaction vector in MnSi has a component along the [0 1 1]
crystallographic direction which induces paramagnetic fluctuations
centered at positions incommensurate with the chemical lattice.

In order to proceed further with the analysis we assume for the transverse
susceptibilities in Eq.~\ref{pa} the expression for itinerant
magnets as given by self-consistent re-normalization theory (SCR)
\cite{moriya85}
\begin{equation}
 \chi^\perp (\vec q \pm \vec \delta, \omega) =
           \chi^\perp (\vec q \pm \vec \delta)/(1-i\omega/\Gamma_{\vec q \pm \vec \delta}).
\label{src}
\end{equation}
$\vec \delta$ is the ordering wave-vector, $\chi^\perp (\vec q \pm
\vec \delta)=\chi^\perp(\vec \mp \delta)/(1+q^2/\kappa^2_\delta)$
the static susceptibility, and $\kappa_\delta$ the inverse
correlation length. For itinerant ferromagnets the damping of the
spin fluctuations is given by $ \Gamma_{\vec q \pm \vec \delta} =
uq (q^2+\kappa^2_\delta)$ with $u = u(\vec \delta)$ reflecting the
damping of the spin fluctuations. Experimentally, it has been
found from previous inelastic neutron scattering measurements that
the damping of the low-energy fluctuations in MnSi is adequately
described using the results of the SCR-theory rather than the
$q^z$ ($z = 2.5$) wave-vector dependence expected for a Heisenberg
magnet \cite{ishikawa82}.

The solid lines of Figs. \ref{Fig1} to \ref{Fig3} show fits of
$({{d^2\sigma}/({d\Omega d\omega}}))_{p}$ to the polarized beam
data. It is seen that the cross section for itinerant magnets
reproduces the data well if the incommensurability is properly
taken into account. Using Eqs. \ref{pa} and \ref{src} and taking
into account the resolution function of the spectrometer, we
extract values $\kappa_0 = 0.12$ \AA$^{-1}$ and $u = 27$
meV\AA$^3$ in reasonable agreement with the analysis given in Ref.
\cite{ishikawa85}. The smaller value for $u$ when compared with $u
= 50$ meV\AA$^3$ from Ref. \cite{ishikawa82} indicates that the
incommensurability $\vec \delta = (0.02,0.02,0.02)$ was neglected
in the analysis of the non-polarized neutron data. At $T = 40$ K,
the chiral fluctuations are broad (Fig. \ref{Fig3}) due to the
increase of $\kappa_\delta$ with increasing $T$, i.e.
$\kappa_\delta(T) = \kappa_0 (1-T_C/T)^\nu$. We note that the
mean-field-like value $\nu = 0.5$ obtained here 
is close to the expected
exponent $\nu = 0.53$ for chiral symmetry \cite{kawamura_88}.
This suggests that 
a chiral-ordering transition also occurs in MnSi in a 
similar way to the rare-earth compound Ho, pointing toward 
the existence of a universality class in the magnetic ordering of
helimagnets~\cite{plakhty_01}.

In conclusion, we have shown that chiral fluctuations can be
measured by means of polarized inelastic neutron scattering in
zero field, when the antisymmetric part of the dynamical
susceptibility has a finite value. We have shown that this is the
case in metallic MnSi that has a non-centrosymmetric crystal
symmetry. For this compound the axial interaction leading to the
polarized part of the neutron cross-section has been identified as
originating from the DM-interaction. Similar investigations can be
performed in a large class of other physical systems. They will
yield direct evidence for the presence of antisymmetric
interactions in forming the magnetic ground-state in magnetic
insulators with DM-interactions, high-T$_c$ superconductors (e.g.
La$_2$CuO$_4$ \cite{berger}), nickelates \cite{koshibae},
quasi-one dimensional antiferromagnets \cite{tsukada} or metallic
compounds like FeGe\cite{lebech}.



%

\begin{figure}
\centering
\caption{ Contour-map of the polarization dependent scattering
  for an energy transfer $\hbar\omega = 0.5$ meV as measured near the (0 1 1) reciprocal lattice point 
at T=35K.
}
\label{Fig4}
\end{figure}


\begin{references}

\bibitem{plakhty_01}V.P. Plakhty et al., Phys. Rev. B {\bf 64}, 100402(R), 2001.
\bibitem{kawamura_88}H. Kawamura, Phys. Rev. B {\bf 38} 4916 (1988).
\bibitem{mason_89} T.E. Mason et al., Phys. Rev. B {\bf 39} 586 (1989).
\bibitem{sulewski} P.E. Sulewski et al., Phys. Rev. Lett. {\bf 67}, 3864 (1991).
\bibitem{maleyev95} S.~V.~Maleyev, Phys. Rev. Lett. {\bf 75}, 4682 (1995).
\bibitem{plakhty99} V.~P.~Plakhty et al., Europhys. Lett. {\bf 48}, 215 (1999).
\bibitem{moriya85} T. Moriya, in \textit{Spin Fluctuations in Itinerant Electron Magnetism}
                   \textbf{56}, Springer-Verlag, Berlin Heidelberg New-York Tokyo, 1985.
\bibitem{ishikawa77}Y. Ishikawa et al., Phys. Rev. B \textbf{16}, 4956 (1977).
\bibitem{ishikawa82}Y. Ishikawa et al., Phys. Rev. B \textbf{25}, 254 (1982).
\bibitem{ishikawa85}Y. Ishikawa et al., Phys. Rev. B \textbf{31}, 5884 (1985).
\bibitem{shirane83} G.~Shirane et al., Phys. Rev. B \textbf{28}, 6251 (1983).
\bibitem{izyumov} e.g. Yu A. Izyumov, Sov. Phys. Usp. \textbf{27}, 845
(1984). 
\bibitem{tixier97} S. Tixier et al., Physica B {\bf 241-243}, 613, (1998).
\bibitem{bloch75} Y. Ishikawa et al., Solid. State. Commun. {\bf 19}, 525 (1976).
\bibitem{remanent} No spin flipping devices are necessary due to the remanent magnetization of the
                   supermirror coatings of the benders. For details see: {P. B\"oni et al., Physica B \textbf{267-268}, (1999) 320.}
\bibitem{semadeni01} F.~Semadeni, B.Roessli, and P.~B\"oni, Physica B \textbf{297}, 152 (2001).
\bibitem{lovesey} S.W. Lovesey and E. Balcar, Physica B {\bf 267-268}, 221 (1999).
\bibitem{zheludev} A. Zheludev et al., Phys. Rev. Lett. \textbf{78}, (1997) 4857.
\bibitem{roessli} B. Roessli et al., Phys. Rev. Lett. \textbf{86} (2001) 1885.
\bibitem{dzyal58} L.~Dzyaloshinskii, J. Phys. Chem. Solids {\bf 4}, 241 (1958).
\bibitem{moriya60} T.~Moriya, Phys. Rev. {\bf 120}, 91 (1960).
\bibitem{kataoka_84}M. Kataoka et al., J. Phys. Soc. Japan {\bf 53}, 3624 (1984).
\bibitem{aristov_00} D.N. Aristov and S.V. Maleyev Phys. Rev. B \textbf{62} {(2000)} R751.
\bibitem{berger}J. Berger and A. Aharony, Phys. Rev. B \textbf{46}, 6477 (1992).
\bibitem{koshibae} W. Koshibae, Y. Ohta and S. Maekawa, Phys. Rev. B \textbf{50}, 3767 (1994).
\bibitem{tsukada} I. Tsukada et al., Phys. Rev. Lett.\textbf{87}, 127203 (2001).
\bibitem{lebech} B. Lebech, J. Bernhard, and T. Freltoft, J. Phys.: Condens. Matter \textbf{1}, 6105 (1989).

\end{references}
\end{document}